\documentclass[10pt,draft]{dis03}
\usepackage{epsf,amsmath}

\newcommand{\msbar}{\mbox{$\overline{\rm{MS}}$}\ }

\begin{document}%
\title{ZEUS results on high \boldmath$Q^2$ DIS cross sections and QCD fits} 
\author{Julian~Rautenberg%
\thanks{Supported by the German Federal Ministry for Education and Research
(BMBF) under contract number HZ1PDA 5}\\
Physikalisches Institut der Universit\"at Bonn\\
D-53115 Bonn, Germany\\
E-mail: Julian.Rautenberg@desy.de }
\maketitle
\begin{abstract}
\noindent 
Single and double differential cross sections 
$d\sigma/dQ^2, d\sigma/dx, d\sigma / dy$ and
$d^2\sigma / dQ^2dx$ 
of the inclusive neutral current (NC) measurement from the 98/99 $e^-p$ 
and the charged current (CC) measurement from the 99/00 $e^+p$ data sets
are presented. 
With the previously published 
$e^+p$ NC ZEUS measurement $xF_3$ and $xG_3$, 
and with the $e^-p$ CC ZEUS measurement $F_2^{CC}$
are extracted.
The results are compared to the
prediction of the Standard Model (SM) 
using parton density functions (PDFs) of
the ZEUS next-to-leading-order (NLO) 
QCD analysis which is reviewed.

\end{abstract}

\paragraph{Introduction}

For longitudinally unpolarized beams
the  deep inelastic scattering (DIS) cross section
for  $e p \rightarrow l' X$ 
can be written
in leading order of the electroweak interaction:\\[1pt]
\centerline{
$\displaystyle\frac{d^{2}\sigma_{\rm{Born}}}{dxdQ^{2}} = A \cdot 
[Y_{+}F_{2}(x,Q^{2})-y^{2}F_{\rm{L}}(x,Q^{2}) - Y_{-}xF_{3}(x,Q^{2})],$} 
where $Q^2$ is the negative square of the four-momentum transfer,
$x$ the Bjorken scaling variable, 
$y$ the inelasticity
and $Y_{\pm}=1\pm (1-y)^2$.
The coefficient $A$ containing couplings and propagator is 
$A^{\rm NC}=\frac{2\pi\alpha^2}{xQ^4}$ for NC 
and $A^{\rm CC}=\frac{G^{2}_{\rm{F}}}{4\pi x} \frac{M_{W}^{4}}{(Q^{2}+M_{W}^{2})^{2}}$
for CC where $G_{\rm{F}}$ is the Fermi constant
and  $M_W$ the $W$ boson mass.
At leading order in QCD 
the structure functions $F_2$ and $xF_3$
may be written in terms of sums and differences of quark 
and antiquark PDFs, 
but for CC depend on the charge of the incoming lepton:\\
\centerline{
$\displaystyle \begin{array}{@{}l@{\;}c@{\;}rcl@{\;}c@{\;}r@{}}
F_2^{{\rm NC}}&=&x\sum_{q=u...b}{A_f\left[q+\bar{q}\right]}&\qquad&  
F_{2,e^{+}p}^{CC} &=& x[d+s+\bar{u}+\bar{c}]\\
F_3^{{\rm NC}}\!&\!=\!&\!x\sum_{q=u...b}{B_f\left[q-\bar{q}\right]}&&
xF_{3,e^{+}p}^{CC} &=& x[d+s-\bar{u}-\bar{c}]\ .
\end{array}$}\\[1pt]
In case of the electromagnetic ($\gamma$) and weak ($Z$) NC processes
the flavor-dependent couplings are included as coefficients
$A_f$ and $B_f$,
while for the purely weak ($W^\pm$) CC process the couplings are global 
and included in $A^{\rm CC}$.
For NC the $t$-quark 
and for CC 
both third generation quarks can be ignored at HERA center-of-mass energies. 
The longitudinal structure function, $F_{\rm{L}}$, 
is zero at leading order QCD and negligible at NLO except
at values of $y$ close to 1.
The reduced cross sections are defined as:\\[4pt]
\centerline{
$\displaystyle\tilde{\sigma}^{\rm NC}=\frac{xQ^4}{2\pi\alpha^2Y_+}\frac{d^2\sigma^{{\rm NC}}}{dxdQ^2}
\qquad\quad
\tilde{\sigma}^{\rm CC}=\frac{2\pi x}{G^{2}_{\rm{F}}} \frac{(Q^{2}+M_{W}^{2})^{2}}{M_{W}^{4}}\frac{d^2\sigma^{{\rm CC}}}{dxdQ^2}\ .%
$}

\paragraph{ZEUS NLO QCD fits} 
The ZEUS data can be used to determine
the input parameters, i.e.   
the PDFs and the value of the strong coupling constant, $\alpha_s$,
of the DGLAP evolution which describes 
DIS measurements over a broad kinematic range.
The ZEUS-S fit~\cite{pr:d67:012007} uses ZEUS data from 
96/97~\cite{epj:c21:443} 
together with fixed-target data to extract PDFs.
Including $\alpha_s$ as a free parameter \mbox{(ZEUS-$\alpha_s$ fit)}
correlations between the PDFs and $\alpha_s$ are fully taken into account,
yielding $\alpha_s(M_Z)=0.1166\pm0.0053 ({\rm tot.})$.
The ZEUS-O fit explores the power of ZEUS data only 
by using ZEUS CC $e^+p$ data from 
94-97~\cite{epj:c12:411}, and the CC and NC $e^-p$
data from the 98/99 runs~\cite{pl:b539:197,desy-02-113}
together with the 96/97 $e^+p$ NC data.
The cut $Q^2\geq 2.5\,{\rm GeV}^2$ 
ensures the applicability of pQCD while the cut $W^2 > 20\,{\rm GeV}^2$
reduces target mass and higher twist sensitivity.
The kinematic range covered by the input data to the fits 
is $6.3 \times 10^{-5} \leq x \leq~0.65$ and 
$ 2.5\leq Q^2 \leq 30\,000\,{\rm GeV}^2$.
The calculations were performed at leading twist 
using the \msbar{}-scheme and the TRVFN treatment of heavy quarks.
The PDFs 
$u$ valence, $xu_v(x)$, $d$ valence, $xd_v(x)$, 
total sea, $xS(x)$,
gluon, $xg(x)$, and the difference between the $d$ and $u$
contributions to the sea, $x\Delta=x(\bar d-\bar u)$,
were parameterized at $Q^2_0 = 7\,{\rm GeV}^2$ by:\\
\centerline{$   xf(x) = p_1 x^{p_2} (1-x)^{p_3}( 1 + p_5 x)$}
Additional constraints, e.g. number or momentum sum rule,
fix 9 of 20 parameters,
given in brackets (without uncertainties) when constrained 
(fixed) below,
leaving 11 free.
The results of the ZEUS-S fit given 
with first (second) statistical and uncorrelated (correlated) 
uncertainties are:\\[2pt]
\centerline{ \scriptsize
\begin{tabular}{|@{\;}l@{\;}|@{\;}c@{\;}|@{\;}c@{\;}|@{\;}c@{\;}|@{\;}c@{\;}|}
\hline
             \scriptsize{}&
             \scriptsize{$p_1$}&
             \scriptsize{$p_2$}&
             \scriptsize{$p_3$}&
             \scriptsize{$p_5$} \\
\hline
             \scriptsize{$xu_v$} &    
             \scriptsize{($1.69 \pm\! 0.01 \pm\! 0.06$)} &
             \scriptsize{$0.5$}                      &
             \scriptsize{$4.00 \pm\! 0.01 \pm\! 0.08$}   &
             \scriptsize{$5.04 \pm\! 0.09 \pm\! 0.64$}   \\
\hline
             \scriptsize{$xd_v$} &
             \scriptsize{($0.96 \pm\! 0.01  \pm\! 0.08$} &
             \scriptsize{$0.5$}                      &
             \scriptsize{$5.33 \pm\! 0.09 \pm\! 0.48$}   &
             \scriptsize{$6.2 \pm\! 0.4 \pm\! 2.3$}      \\
\hline
             \scriptsize{$xS$} &
             \scriptsize{$0.603 \pm\! 0.007 \pm\! 0.048$} &
             \scriptsize{$-0.235 \pm\! 0.002 \pm\! 0.012$}&
             \scriptsize{$ 8.9 \pm\! 0.2 \pm\! 1.2$}     &
             \scriptsize{$6.8 \pm\! 0.4 \pm\! 2.0$}      \\
\hline
             \scriptsize{$xg$}  &
             \scriptsize{($1.77 \pm\! 0.09 \pm\! 0.49$)} &
             \scriptsize{$-0.20 \pm\! 0.01 \pm\! 0.04$}  &
             \scriptsize{$6.2 \pm\! 0.2 \pm\! 1.2$}      &
             \scriptsize{$0$}                        \\
\hline
             \scriptsize{$x\Delta$} &
             \scriptsize{$0.27 \pm\! 0.01 \pm\! 0.06$}  &
             \scriptsize{$0.5$}                     &
             \scriptsize{($10.9 \pm\! 0.2 \pm\! 1.2$)}  &
             \scriptsize{$0$}                       \\
\hline
\end{tabular}}\\[4pt]
\begin{figure}[!tb]{%
\label{pdfs}
\begin{tabular}[c]{@{}c@{}c@{\;}c@{}}
\epsfxsize=0.32\textwidth\epsfbox{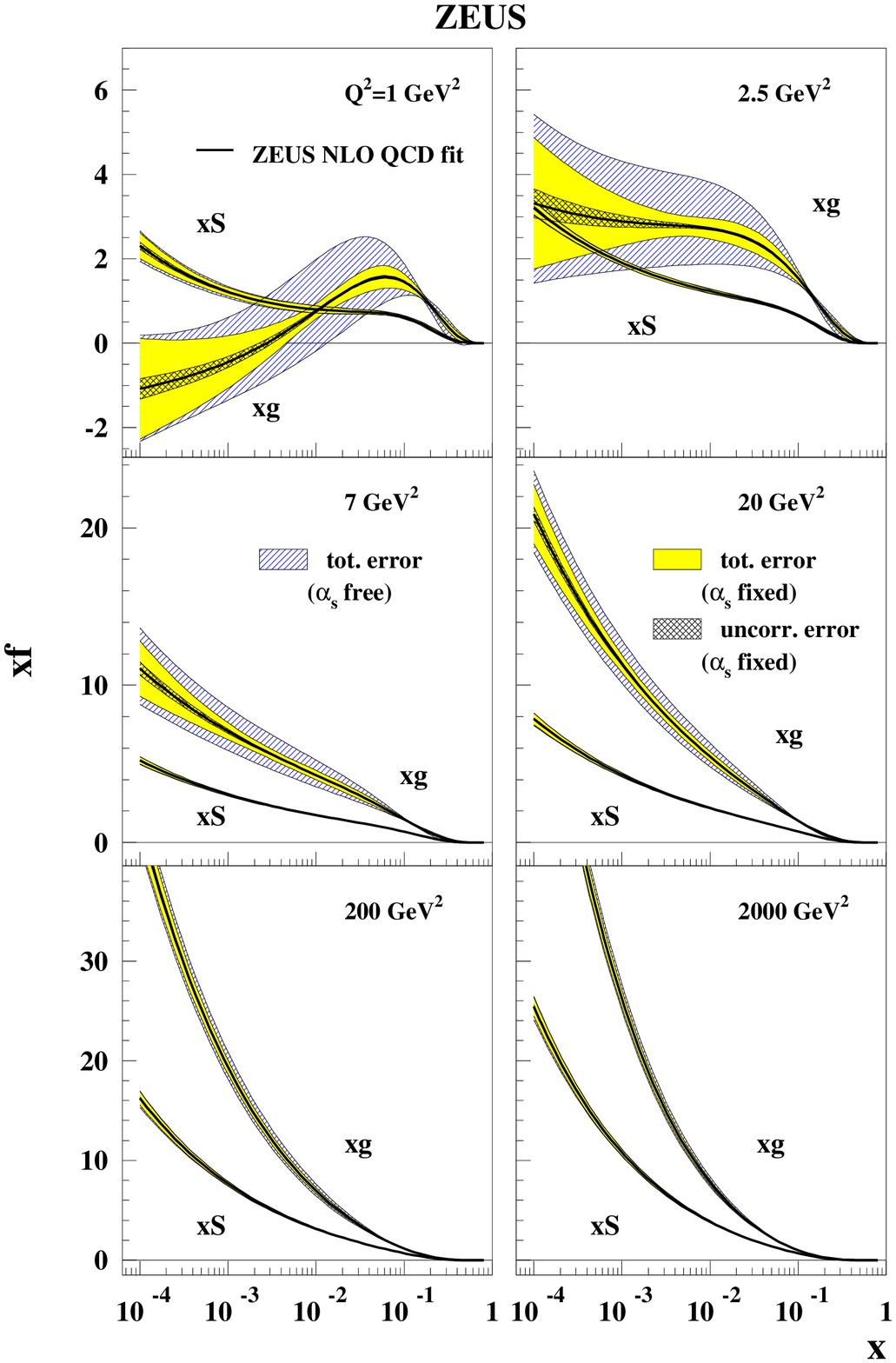}&
\epsfxsize=0.32\textwidth\epsfbox{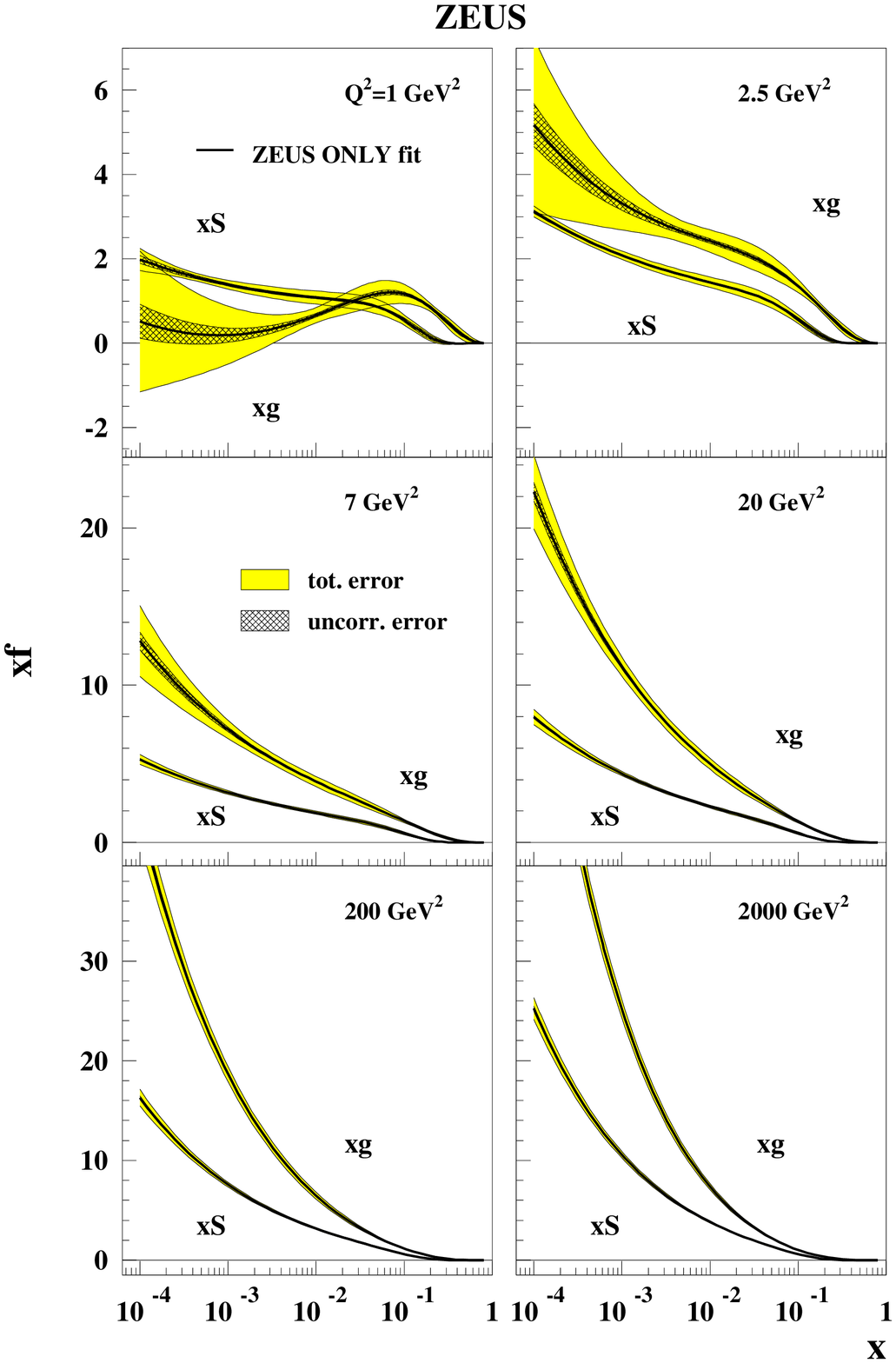}&
\parbox[b]{0.35\textwidth}{%
\caption[*]
{Gluon and sea distributions from  
ZEUS-S (a) and ZEUS-O (b)
for various $Q^2$ values with
statistical and uncorrelated systematic uncertainty (cross-hatched)
including correlated systematic uncertainties (grey)
and in (a) the total experimental uncertainty from ZEUS-$\alpha_s$ 
(hatched).}}
\end{tabular}}\end{figure}
In Fig.~1 $xg(x)$ and $xS(x)$ are
displayed for the ZEUS-S and ZEUS-O results with uncertainty bands,
where the correlated systematic uncertainties are dominant.
The gluon density is found not to rise towards low $x$ 
as dramatically as expected for low $Q^2$.
For ZEUS-S it even becomes negative.
Since $xg(x)$ is not an observable like a structure function
this is not unphysical, but might indicate that NLO-pQCD is inadequate.
While uncertainties for the ZEUS-O fit are considerable larger 
than for ZEUS-S, a preliminary fit result including in addition 
the high luminosity data from 99/00 yields
compatible uncertainties without including the fixed-target data.

\paragraph{\boldmath $e^-p$ NC measurement}
\label{sec-nc}

\begin{figure}[!p]{%
\label{nc_dsqxy}
\begin{center}
\begin{tabular}[c]{@{}c@{\;}c@{}}
\epsfxsize=0.456\textwidth\epsfbox{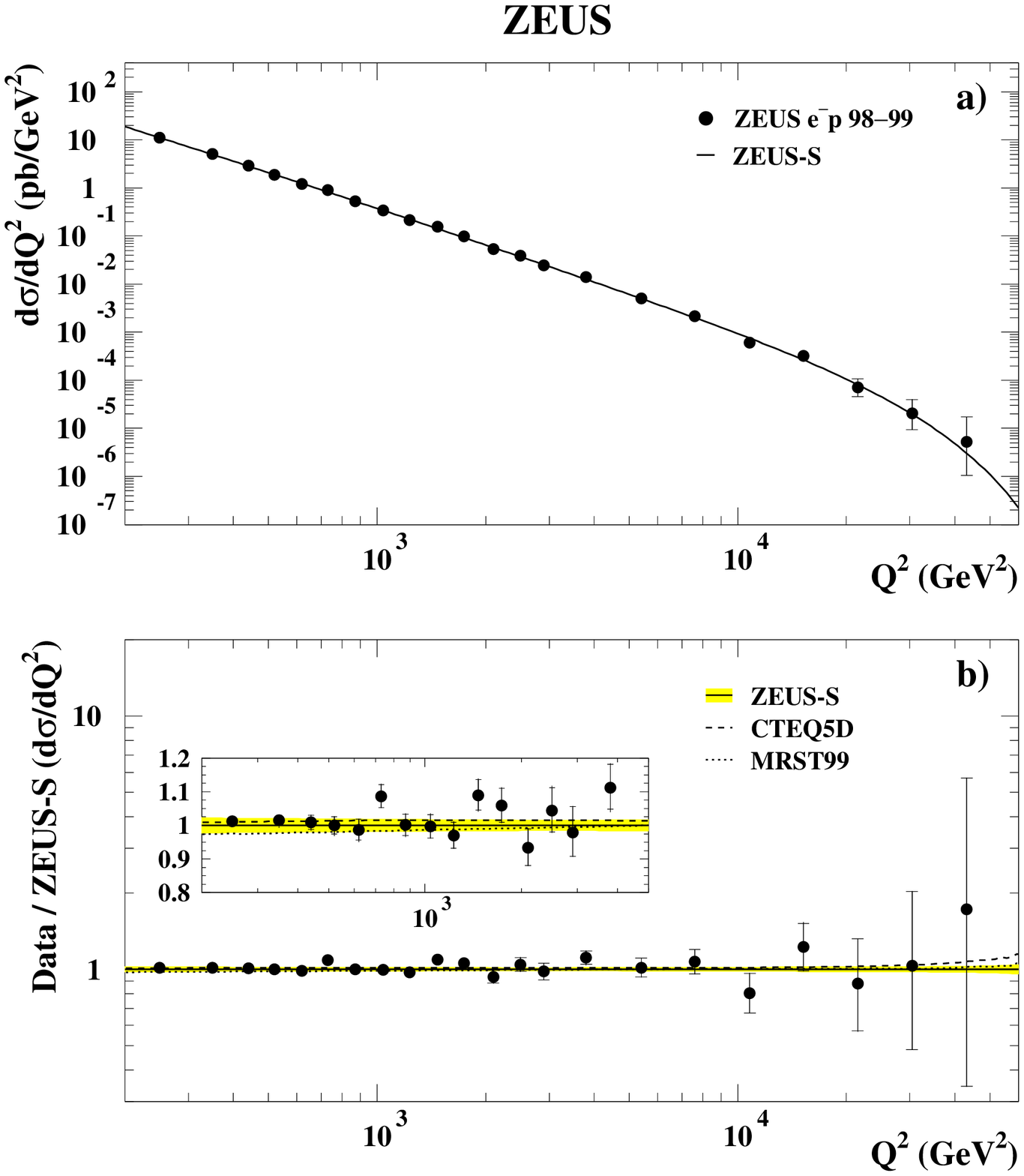}&
\epsfxsize=0.532\textwidth\epsfbox{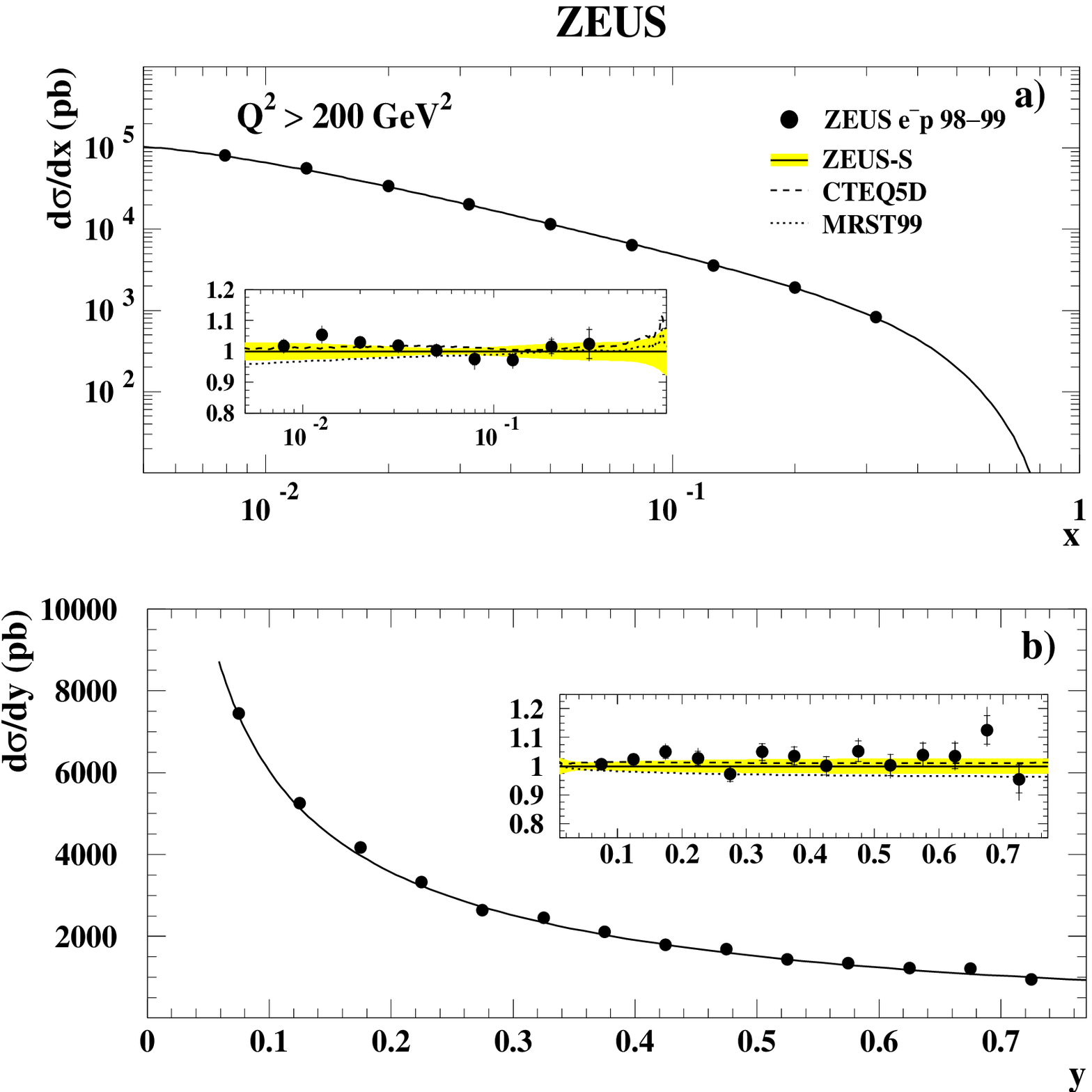}
\end{tabular}
\caption[*]
{NC cross sections $d\sigma/dQ^2,d\sigma/dx$ and $d\sigma/dy$ 
and ratio to SM using ZEUS-S PDFs.}
\end{center}}
\end{figure}
\begin{figure}[!p]{%
\label{nc_rs_xFG}
\begin{tabular}[c]{@{}c@{\;}c@{}}
\epsfxsize=0.515\textwidth\epsfbox{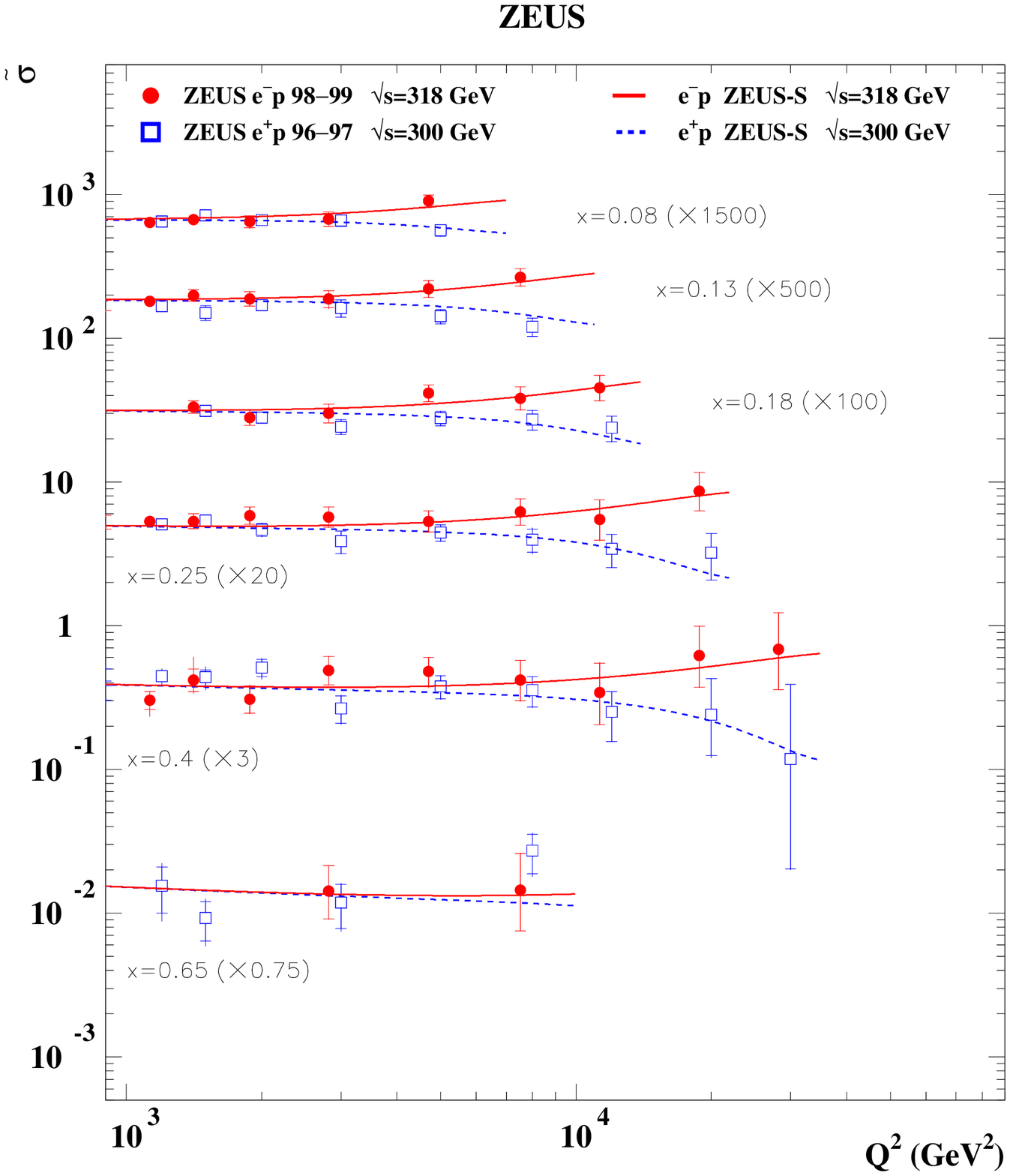}&
\parbox[b]{0.476\textwidth}{
\epsfxsize=0.475\textwidth\epsfbox{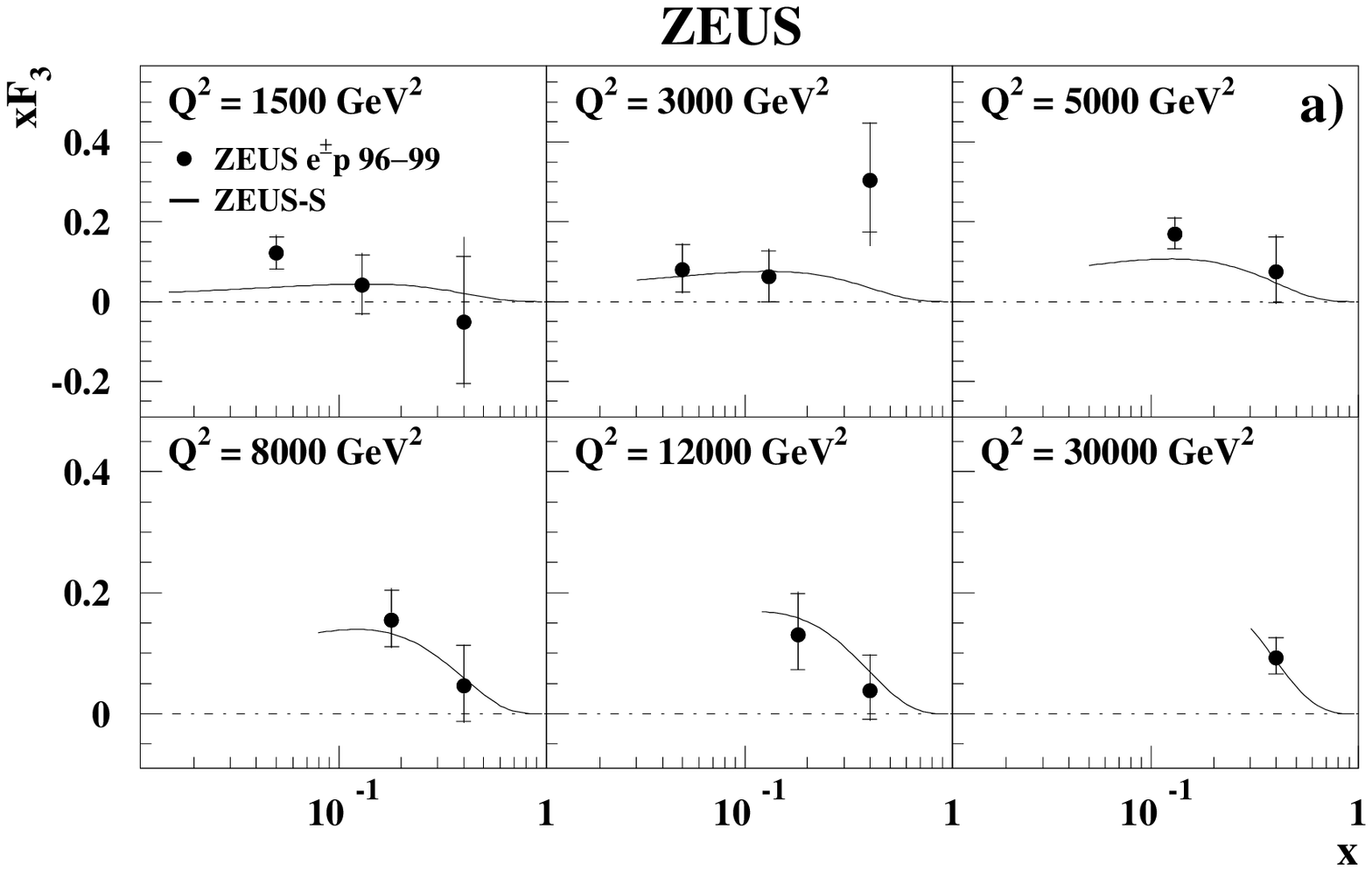}
\epsfxsize=0.475\textwidth\epsfbox{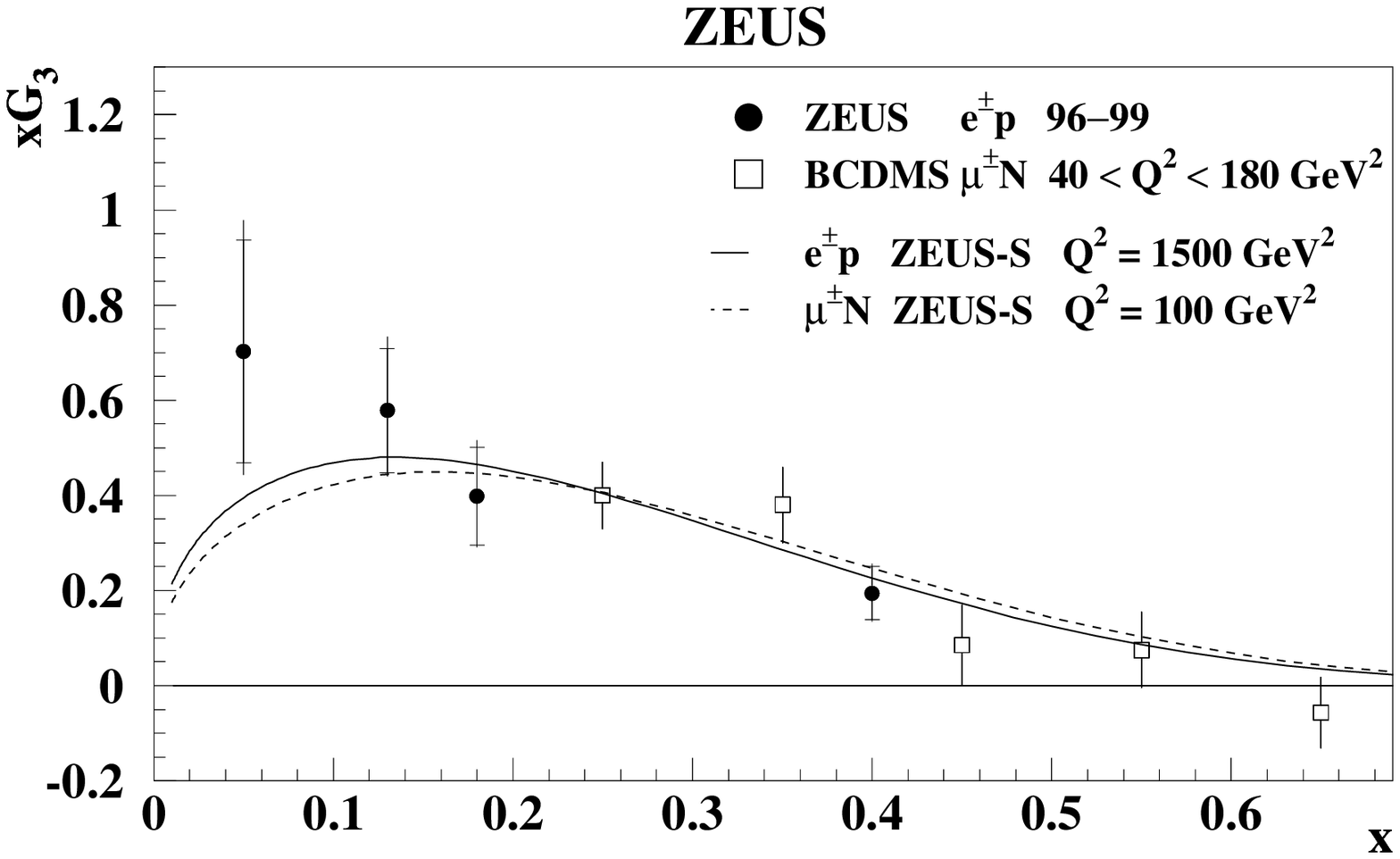}}
\end{tabular}
\caption[*]
{NC reduced cross section $\tilde{\sigma}^{\rm NC}$ for $e^-p$ 
and $e^+p$ data sets, and the extracted $xF_3$ and $xG_3$.}}
\end{figure}
The measurement~\cite{desy-02-113} is based on the $e^-p$ 98/99 running period 
coresponding to an integrated luminosity of  16 pb$^{-1}$
in the kinematic region $185\,{\rm GeV}^2 < Q^2 < 50\,000\,{\rm GeV}^2$ and 
$0.0037 < x < 0.75$.
The kinematic variables are reconstructed using the double angle method.
The event selection is based on identifying the scattered electron
fullfilling criteria on isolation, track matching, 
energy and transverse momentum.
Background is suppressed by requiring 
a proper event vertex,
transverse ($P_T/\sqrt{E_T}<4\sqrt{\rm GeV}$) and
longitudinal ($38\,{\rm GeV}<E-p_z<65\,{\rm GeV}$) momentum conservation.
Restricting $y$ ($y_E<0.95$) excludes photoproduction.
The measured cross sections in Fig.~2 show excellent agreement 
with the SM using ZEUS-S.
Comparing the $e^-p$ to the $e^+p$ reduced cross sections in Fig.~3
the parity violating effect of the weak $Z$ exchange 
(and $\gamma$-$Z$ interference)
becomes visible at high $Q^2$. 
From the difference, $xF_3$ is extracted, shown in the same figure.
Separating $xF_3$ in interference 
and $Z$ exchange terms yields structure functions $xG_3$ and $xH_3$:\\
\centerline{$xF_3=-a_e\chi_ZxG_3 + 2a_ev_e\chi_Z^2xH_3, \quad 
\chi_Z= \kappa_W\cdot{Q^2}/({M_Z^2+Q^2}).$}
The second term can be neglected to reconstruct $xG_3$ which extends the
low $Q^2$ muon-carbon measurements from BCDMS to higher $Q^2$, as shown in Fig.~3.

\paragraph{\boldmath $e^+p$ CC measurement}
\label{sec-cc}
\begin{figure}[!t]{%
\label{cc_dsqxy}
\begin{tabular}[c]{@{}c@{\;}c@{\;}c@{}}
\epsfxsize=0.32\textwidth\epsfbox{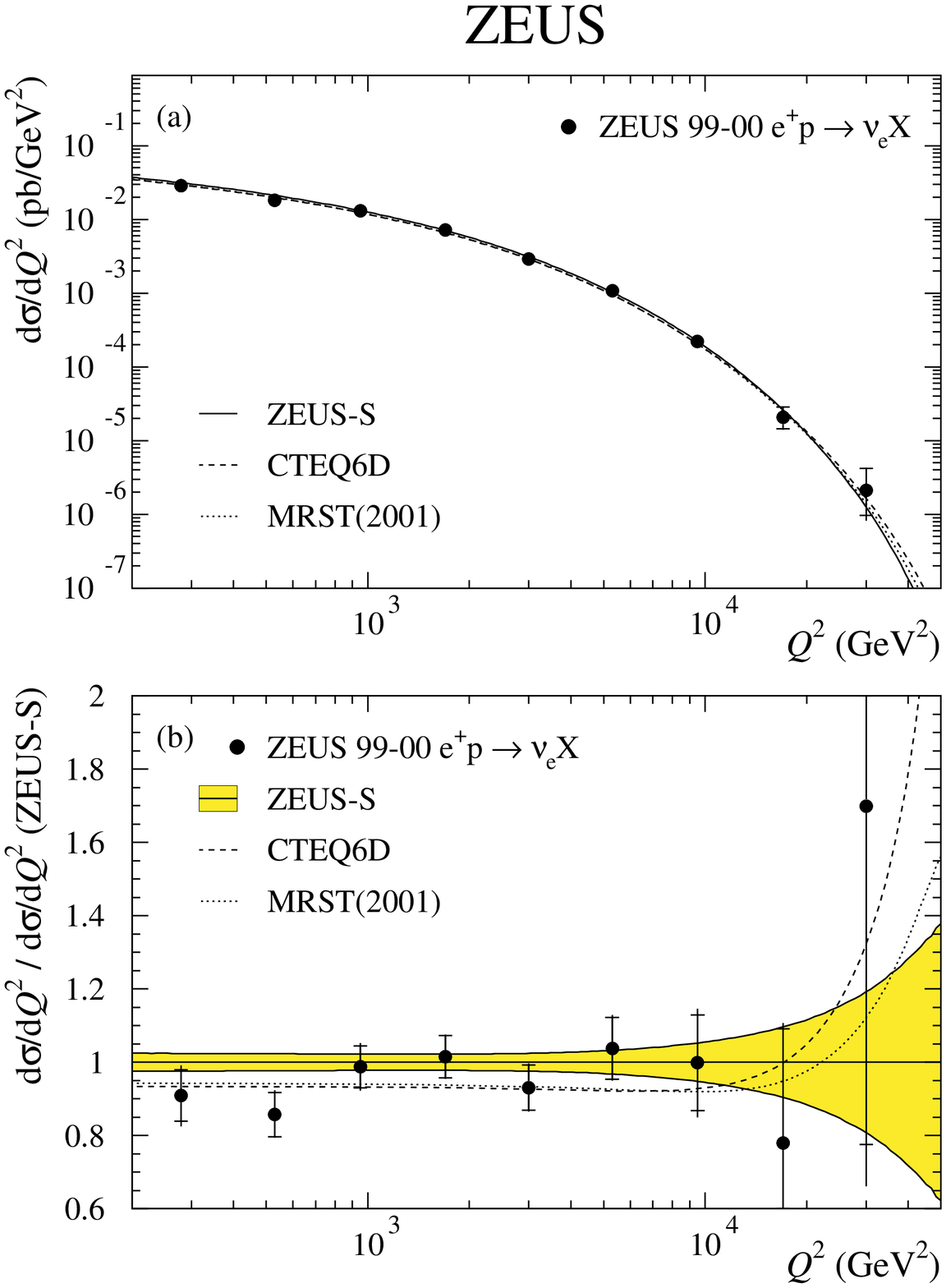}&
\epsfxsize=0.32\textwidth\epsfbox{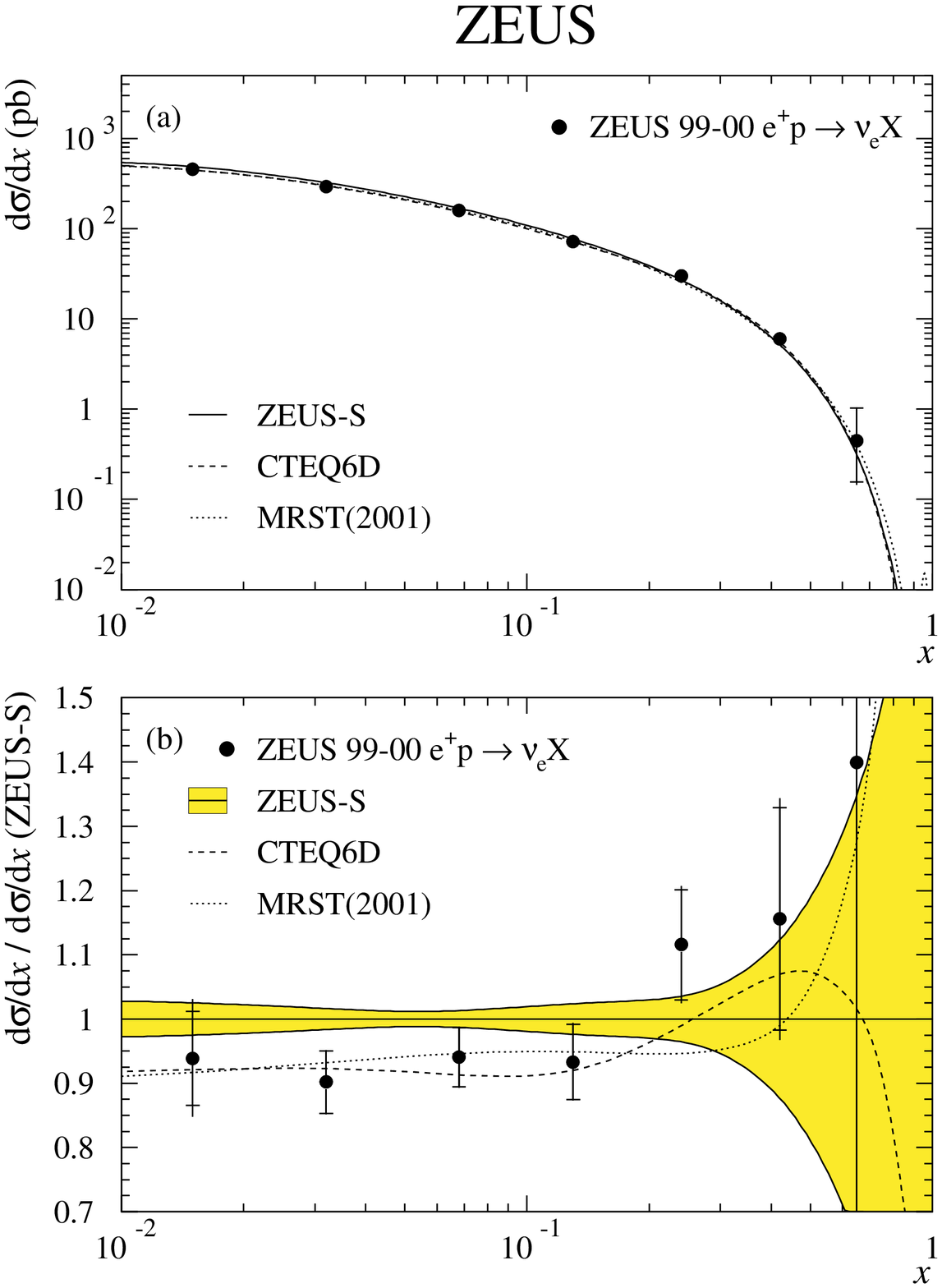}&
\epsfxsize=0.32\textwidth\epsfbox{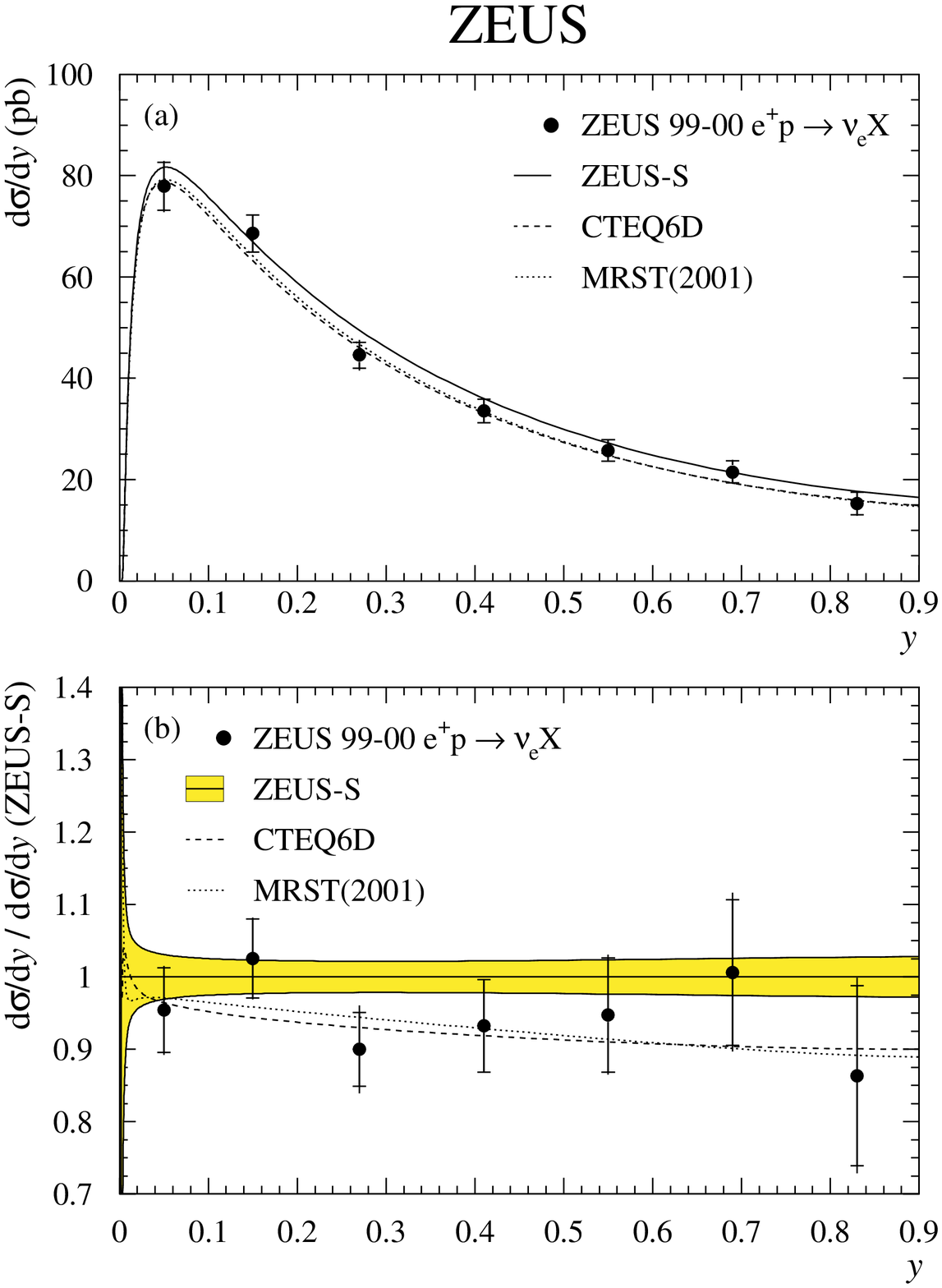}
\end{tabular}
\caption[*]
{CC cross sections $d\sigma/dQ^2,d\sigma/dx$ and $d\sigma/dy$
and ratio to SM using ZEUS-S PDFs.}
}\end{figure}
The CC measurements~\cite{desy-03-093}
are based on the $e^+p$ data from 99/00 
corresponding to an integrated luminosity of 61 pb$^{-1}$.
CC events are selected by missing transverse momentum 
due to the undetected neutrino
and by vetoing events with a 
candidate scattered electron.
For a sample of large hadronic angle $\gamma_{\rm had}\!\!>\!\!0.4\,{\rm rad}$ 
the acceptance of the central tracking detector allows for efficient 
background suppression, while for a sample of low $\gamma_{\rm had}$
the cuts on the calorimetric quantities are raised.
The excellent agreement of the measured cross sections with the SM using ZEUS-S 
is shown in Fig.~4. 
Using the previously published $e^-p$ CC ZEUS results~\cite{pl:b539:197}
the singlet structure function $F_2^{\rm CC}$ is extracted and extends the CCFR measurement 
in Fig.~5 
by two orders towards higher $Q^2$.
The dependence of the cross section on the propagator mass  
is used to extract $M_W$ in the space-like region in a fit to $d\sigma^{\rm CC}/dQ^2$:\\
\centerline{$M_W=\left( 78.9\pm2.0 ({\rm stat.}) 
  \pm{-1.8} ({\rm syst.}) 
  ^{+2.0}_{-1.8} ({\rm PDF})\right) \mbox{GeV}$,}
This is in good agreement with the more precise time-like measurements giving
 $M_W=80.422\pm0.047\,{\rm GeV}$ \cite{epj:c15:1}.
\begin{figure}[!t]{%
\label{cc_rs_f2}%
\begin{tabular}[c]{@{}c@{\;}c@{\;}c@{}}
\epsfxsize=0.39\textwidth\epsfbox{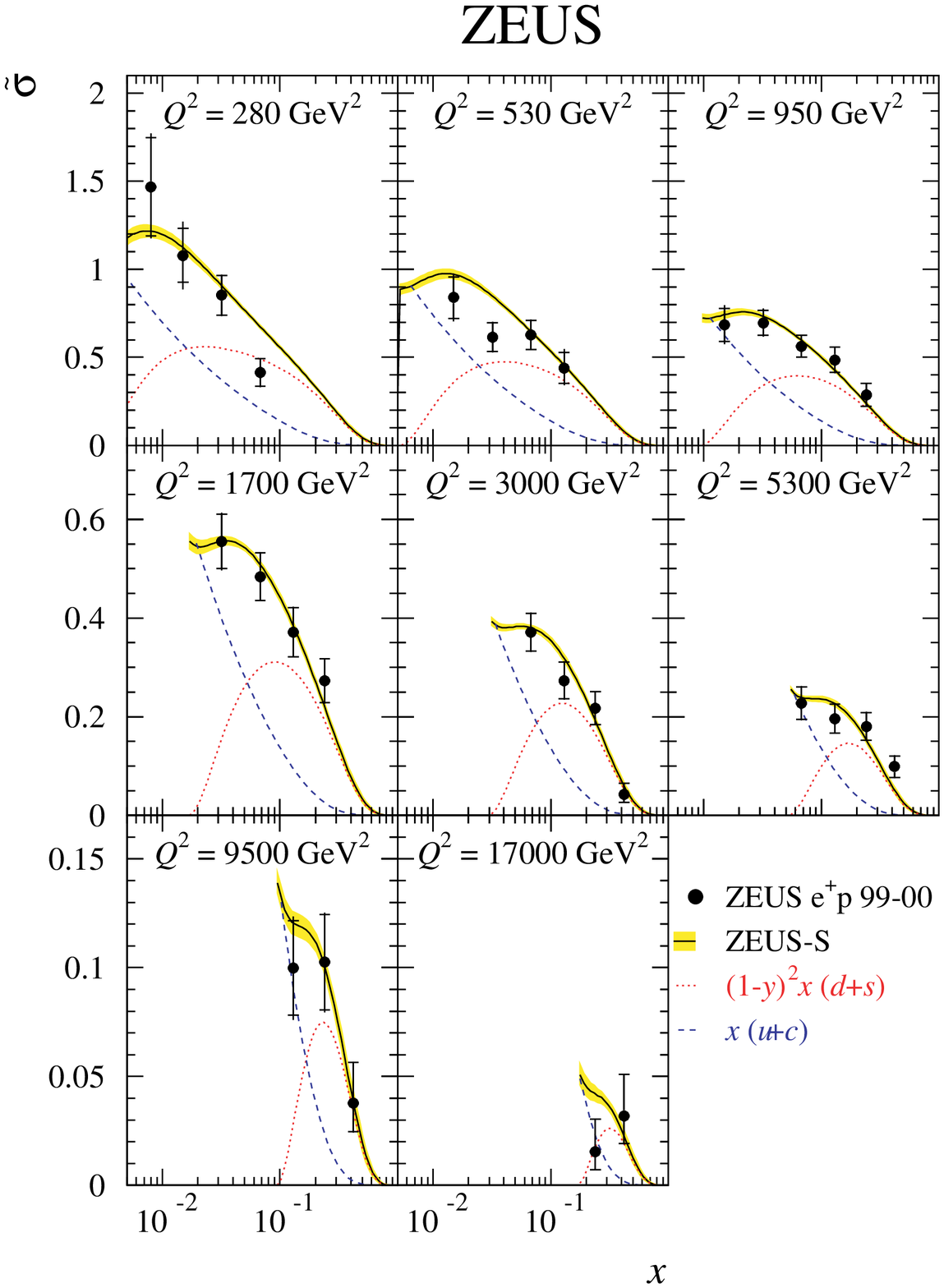}&
\epsfxsize=0.39\textwidth\epsfbox{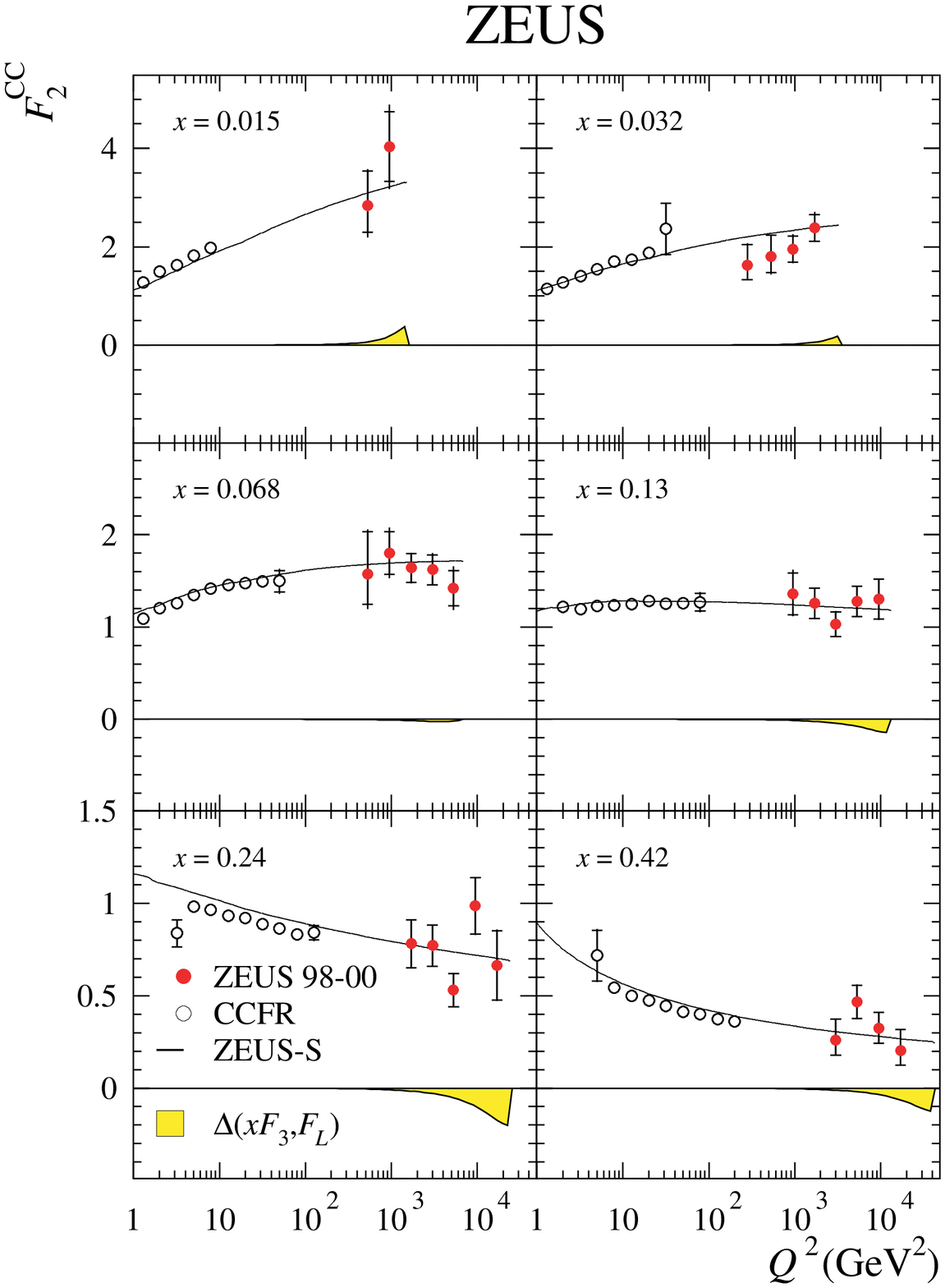}&
\parbox[b]{0.2\textwidth}{%
\caption[*]%
{Reduced CC cross sections 
$\tilde{\sigma}^{\rm CC}$ for fixed values of $Q^2$ (left),
and $F_2^{\rm CC}$ for fixed values of $x$ together with BCDMS
 data (right), both with SM prediction using ZEUS-S PDFs.}}
\end{tabular}}\end{figure}

\paragraph{Summary}
\label{sec-sum}
The inclusive NC measurement from the 98/99 $e^-p$ 
data set coresponding to an integrated luminosity of 16 pb$^{-1}$ 
and the CC measurement from the 99/00 $e^+p$ data set
 of 61 pb$^{-1}$ 
are presented. 
With the
$e^+p$ NC ZEUS measurement $xF_3$ and $xG_3$, 
and with the $e^-p$ CC ZEUS measurement $F_2^{CC}$
are extracted.
All measurements are in excellent agreement with the 
SM prediction using the ZEUS-S PDFs from the
 ZEUS NLO QCD analysis.

\end{document}